\documentstyle[twocolumn,prb,aps,epsfig]{revtex}

\begin{document}

\title{A microscopic model for  a class of mixed-spin
quantum antiferromagnets }

\author{J.V. Alvarez$^{1}$ , Roser Valent\'{\i}$^{1}$, A.
Zheludev$^{2,*}$
}
\address{$^1$Fakult\"at 7, Theoretische Physik,
 University of the Saarland,
66041 Saarbr\"ucken, Germany.\\
        $^2$ Solid State Division, Oak Ridge National Laboratory,
Oak Ridge, TN  37831-6393, USA.}

\maketitle

\begin{abstract}

  We propose a microscopic model that describes the magnetic
  beha\-vior of 
  the mixed-spin quantum systems R$_2$BaNiO$_5$ (R= magnetic rare earth).
  An evaluation of the properties of this model by 
  Quantum Monte Carlo simulations shows  remarkable good agreement
 with the experimental data and provides new insight into the
 physics of mixed-spin quantum magnets.
\end{abstract}


\author{J.V. Alvarez}

\author{ Roser Valent\'{\i}}

\author{A. Zheludev}

\maketitle

\vspace*{0.5cm}

Low-dimensional (low-D) quantum magnetism remains at the forefront
of condensed matter research for almost two decades. This is
primarily because this field deals with {\it simple} models of
magnetism that de\-monstra\-te a broad spectrum of {\it complex}
quantum-mechanical phenomena of a general and fundamental nature.
The simplicity of the Hamiltonians involved, as well as the low
dimensionality, often allow accurate theoretical and numerical
treatment. On the other hand, the discovery of real low-D
materials allows experimental studies for direct comparison with
theory on the quantitative level. A particular area of great
recent interest is the co-existence of ``quantum'' and
``classical'' properties in systems composed of weakly coupled
quantum spin chains. One good example is the interplay between
continuum and spin-wave dynamics in gapless quasi-one-dimensional
(quasi-1D) Heisenberg systems such as KCuF$_3$ \cite{KCUF3} and
BaCu$_2$Si$_2$O$_7$ \cite{BACU2}. At the focus of the present work
is a more complex phenomenon: the seemingly paradoxical
co-existence of 3D magnetic long-range order and 1D quantum gap
excitations in rare earth nickelates with the general formula
R$_2$BaNiO$_5$ (R= magnetic rare earth)\cite{Garcia_95,ZHELUDEV98}.

R$_2$BaNiO$_5$ species have two types of spin carriers. $S=1$
Ni$^{2+}$ ions form distinct antiferromagnetic chains running
along the $a$ axis of the crystal structure. If these chains were
perfectly isolated, they would be a classic example of a Haldane
antiferromagnet with no long-range order even at $T=0$ and a gap
in the magnetic excitation spectrum \cite{HALDANE83}. Long-range
ordering that occurs in R$_2$BaNiO$_5$ at low temperatures is
driven by the magnetic $s=\frac{1}{2}$ R$^{3+}$ ions, positioned
in-between the chains. 3D magnetic order involves {\it both} the
R$^{3+}$ and Ni$^{2+}$ spins. What is particularly interesting
though, is that Haldane-gap excitations associated with the
Ni-chains {\it persist} in the ordered state and {\it co-exist}
with conventional order-parameter excitations (spin waves). By
now, a wealth of experimental data on several R$_2$BaNiO$_5$
compounds has been accumulated \cite{Garcia_95,ZHELUDEV98,RAYMOND99}. Most
theoretical work, however, was based on a simple Mean Field (MF)
\cite{ZHELUDEV98} or Random Phase Approximation
(RPA)\cite{ZHELUDEV00} treatment of the interactions between
Ni-chain and rare earth subsystems.   The existing numerical
studies were aimed at calculating the ``bare'' properties of the
Ni-chains, needed to complete the MF/RPA equations\cite{YU99}.
While the MF/RPA approach turned out to be an extremely useful
tool in understanding the basic physics involved, it has numerous
intrinsic limitations, particularly for dilute
(R$_{1-x}$Y$_x$)$_2$BaNiO$_5$ systems \cite{YOKOO98}, or in the
case of strong Ni-R coupling. In the present work we abo\-lish the
MF/RPA framework and construct a {\it microscopic} model for
R$_2$BaNiO$_5$ compounds. We then employ this model in a
systematic first-principle numerical study of these interesting
materials.

  In order to construct the microscopic model we will use the
available experimental results to design an appropriate
description for the magnetic ions and interactions in the system:
(i) A good understanding of the Ni-subsystem can be drawn from the
known properties of Y$_2$BaNiO$_{5}$, a material where the
magnetic rare earths have been replaced by non-magnetic Y$^{3+}$.
The Yttrium-nickelate is an almost perfect physical realization of
a Haldane gap antiferromagnet. Neutron scattering data\cite{YNABO}
show a large gap $\Delta \sim 10$meV and well characterized
Haldane excitations. This  suggests that the intrachain Ni-Ni
exchange coupling in the chains running along the $a$-axis has to
be $J \sim 25$meV. Furthermore, there is no evidence of 3D long
range order even at the lowest temperatures. Any residual {\em
direct} coupling between the chains, both in $b$ or $c$ direction,
is smaller than the critical value necessary to establish AF order
at low enough temperatures \cite{AFFLECK89,SAKAI90}. This critical
value depends quadratically  on the gap and  it is expected to be
large in this family of nickelates due to the magnitude of the
gap. Therefore we will consider the  {\em direct}  Ni-Ni
interchain coupling irrelevant for a physical description of the
material.

(ii) When Y, located between the chains, is completely substituted
by a magnetic rare earth R,  the system orders
antiferromagnetically. The  N\'eel temperature is ty\-pically
smaller than the gap energy.  For example, in the case R=Nd the
$T_N=48$K and $\Delta \sim 11$meV ($\sim 127$K)\cite{ZHELUDEV96}.
In the {\it paramagnetic} phase ($T>T_{\rm N}$), the Ni-chain
excitations in R$_2$BaNiO$_5$ are virtually indistinguishable from
those in the Yttrium compound at the same temperature, implying
that very little should change in the model of the $S=1$ chains
while the ordering should  be driven by some kind of exchange Ni-R
that  induces an {\em indirect} coupling between the chains.

(iii) In the magnetically ordered phase, the staggered
magnetization of the Ni sublattice has  as $T\rightarrow 0$
saturation values of $1.0-1.6\mu_B$ per ion in all the
R$_{2}$BaNiO$_{5}$ studied to date
\cite{ZHELUDEV98,ALONSO90,GARCIA-MATRES93}. These values are
clearly smaller than the classical result $2\mu_B$. This
observation signals that the model describing these compounds has
to properly retain the quantum fluctuations in the ordered phase.

(iv) Unlike Ni-R interactions, that are vital in order to induce a
static magnetization on the Ni sites below $T_N$, direct R-R
magnetic coupling can be disregarded in our model. Indeed, the
ordering temperatures in R$_{2}$BaNiO$_{5}$ compounds are
typically several tens of Kelvin. Direct coupling strength between
rare earths in insulators is much weaker, usually of the order of
one Kelvin. This is due to the fact that in rare earth species the
magnetic $f$-electrons are strongly localized, what prevents
efficient superexchange coupling\cite{HUFNER78}. 
Unlike the case of  Y$_{2-x}$Ca$_{x}$BaNiO$_{5}$ 
\cite{DiTusa_94,DAGOTTO,BATISTA}, where doping with Ca introduces
 hole carriers in the Ni-chains,  this localized 
nature of f-electrons in the rare-earth 
keeps the  S=1 chains free from charge carriers.

(v) In the crystal structure of R$_{2}$BaNiO$_{5}$, the
site-symmetry for R$^{3+}$ is very low. As a result, for each rare
earth ion the degeneracy of its magnetic multiplet is lifted by
crystal field effects. To investigate the low-energy part of the
spectrum we only need to consider the lowest-energy orbital
levels. In the case of the rare earth being a Kramers ion with
half-integer total angular momentum\cite{HUFNER78}, we shall model
the magnetic rare earths by effective $s=\frac{1}{2}$
(pseudo)spins. In order to decide about the type of coupling between the
 Ni and the R sublattices, we rely on the experimental
 observation of
    the absence of dispersion in the crystal field
    excitations associated with R$^{3+}$ ions\cite{ZHELUDEV00}. 
  This fact
 suggests that the coupling  between the R and Ni
sublattices is very anisotropic and can be approximated by an
Ising-type term in the Hamiltonian.

\begin{figure}
 \includegraphics[width=0.35\textwidth]{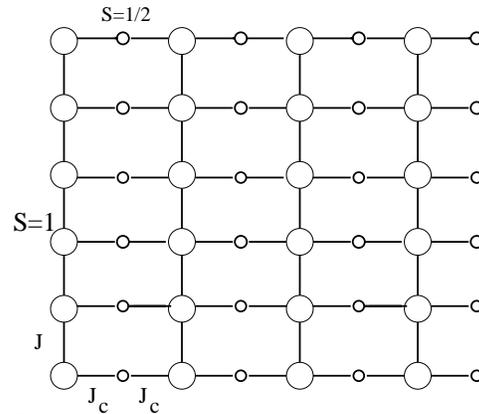}
 \caption{\label{model} Microscopic model for  R$_2$BaNiO$_5$. The
big circles denote $S=1$ Ni spins and the small circles denote the
$s=\frac{1}{2}$ R spins. J and J$_c$ are the Ni-Ni and the Ni-R
exchange couplings respectively.}
 \end{figure}

With all the ingredients above, we propose now the following
Hamiltonian:
         \begin{eqnarray}
        H=J \sum_{ij}{\bf S}_{i, 2j}{\bf S}_{i+1, 2j}+
         J_{c}\sum_{ij} S^{z}_{i,
           2j}(s^{z}_{i, 2j-1}+s^{z}_{i, 2j+1})\nonumber
         \end{eqnarray}
with J and J$_c$ both positive. The first term in $H$
 is  a  $S=1$ Heisenberg model along the Ni
chains, the second is an  Ising-like coupling with strength J$_c$
between the $S=1$ Ni ions and the $s=\frac{1}{2}$ magnetic moment
in the rare earth.  The index $i$ runs along the chain direction
and $j$ in the direction perpendicular to the chains. A sketch of
the model is shown in Fig.\ (\ref{model}). When $J_{c}=0$,
$H$ should reproduce the physics of the
Y$_2$BaNiO$_{5}$, i.e. basically independent $S=1$ chains.

       We have studied this model numerically with  the Quantum
Monte Carlo (QMC) method using a quantum cluster algorithm, the
Loop Algorithm \cite{EVERTZ}. This method allows for an efficient
sampling of the configuration space giving a specific prescription
of how to perform global updates in quantum systems.  Such updates
involve clusters of spins with a size of the correlation length.
The hamiltonian we studied is not frustrated  and does not show the
sign problem.   All the Boltzmann weights can be taken positive 
after the conventional rotation around the z-axis of all the spins 
in one of the two sublattices of the S=1 system. 
The difficulty in reaching extremely low temperatures is a
drawback of the QMC method  when one is interested in characterizing 
exotic ground states, mostly 
near quantum critical points or strongly frustrated systems.
None of these cases affects our calculations since 
 our main goal is to compare with 
experimental results at temperatures, as we will see, 
perfectly accesible for QMC.     
We have measured various  magnitudes that can be directly compared
with the available experimental data and which,  in some cases,
are not accesible for other approaches like the MF/RPA
approximation\cite{ZHELUDEV98,ZHELUDEV00} or the zero temperature
Density Matrix Renormalization Group (DMRG)
calculations\cite{YU99}. These magnitudes are: (i) Magnetizations
in both sublattices (ii) N\'eel temperatures   (iii) Correlation
lengths and gaps in both subsystems.

\vspace*{-0.1cm}


 Before discussing the results, we present  a brief technical
description of the numerical work and data treatment. 
QMC simulations are performed in finite lattices therefore
an exhaustive finite size analysis of the results 
is necessary to compare with real experiments. We used periodic
boundary conditions.
The
magnetization ($M$) was computed  by directly measuring $|M|$ in
order to avoid the averaging between configurations around the
two ground states of the model in the ordered phase,  and by using
the relation $M=\lim_{L \rightarrow \infty }|M|$.  We considered
for this extrapolation L$\times$L lattices of size
L=8,16,24,32,48,64 spins. The N\'eel temperature was then
determined by using the Binder parameter\cite{BINDER00} which is
the fourth cumulant of the order parameter distribution. At $T_N$
this cumulant  is independent of the size of the system, apart
from subdominant corrections. We have observed that for values of
$J_c$ smaller than $\sim 0.1J$ these corrections are  due to the
anisotropy in the scaling (i.e. different number of spins in $a$
and $c$ directions). We have checked this point by studying
systems with different sizes $L_a=32,48$ spins and $L_c=24,32,48$
spins. For values of $J_c \sim 0.3 $ and the precision needed for
comparison with the experiment,  $T_N$ can be computed by
extrapolation from isotropic lattices. The statistical error bars
are always smaller than the symbol sizes in all the figures
presented. We measured all the magnitudes in $10^5$ Monte Carlo
steps after thermalization. We have used units
$\mu_{B}=k_{B}=\hbar=1$.

\begin{figure}
 \includegraphics[width=0.35\textwidth]{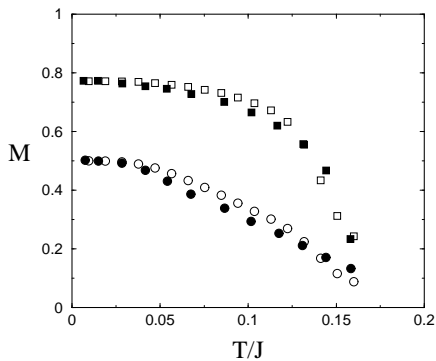}
 \caption{\label{M-T} $M_{\rm Ni}$ (squares) and $M_{\rm R}$
(circles) vs. T. Open symbols are the results obtained with the
model Hamiltonian for $J_c$=0.31$J$ and
filled symbols are the experimental data. Both magnetizations are
given in units of $\mu_B$ and have been rescaled with respect to
the saturation values $1$ and $0.5$ respectively. }
 \end{figure}

The staggered magnetizations obtained from the QMC computations
for  both the  Nickel  ($M_{\rm Ni}$) and rare earth  ($M_{\rm
R}$)  sublattices are in good agreement with the experimental
results for  Nd$_{2}$BaNiO$_{5}$ \cite{ZHELUDEV98} for a suitable
choice of the model parameters in the Hamiltonian. For
instance, for a value of the transverse coupling  $J_c=0.31J$, we have that
$M_{\rm Ni}(T \rightarrow 0)=0.79$ that corresponds to the value
1.6$\mu_B$ observed experimentally. The N\'eel temperature for
this value of $J_c$ is $T_N=0.163J$. We can take the gap in a
single $S=1$ chain, $\Delta=0.410J$ and compute  the ratio between
the two main energy scales in the system
$r=\frac{\Delta}{T_{N}}$.  We observe that $r$ is very similar in
the material ($r$=2.59) and in the model ($r$=2.51). In Fig.\
(\ref{M-T}) we present the temperature dependence of the staggered
magnetizations in the $S=1$ (open squares) and the $s=\frac{1}{2}$
(open circles) sublattices,  $M_{\rm Ni}$ and $M_{\rm R}$
respectively, for the selected value of $J_c$ obtained from the
QMC simulations. The magnetization results have been rescaled with
respect to the maximum saturation values, i.e. $1$ for the $S=1$
system and $0.5$ for the $s=\frac{1}{2}$ system. The results can
be compared with the experimental magnetic moment in the Ni
(filled squares) and Nd systems (filled circles)\cite{ZHELUDEV98}.
Note the good agreement between the theoretical and experimental
results at low temperatures. We also observe that the staggered
magnetization in the $S=1$ sublattice is nearly temperature
independent for $T \leq \frac{T_N}{2}$. At T=40K, the
Nd$_{2}$NiBaO$_{5}$ undergoes a magnetic structural transition
\cite{GARCIA-MATRES97} and therefore  we expect only qualitative
agreement with our model at temperatures near $T_N$.


Both sublattices behave qualitatively different in the ordered
phase. While in the Ni sublattice the reduction of the staggered
magnetization by purely quantum fluctuations is important, the R
sublattice is fully saturated to the classical value $M_{\rm
R}=0.5$ at T=0.

\begin{figure}

 \includegraphics[width=0.4\textwidth]{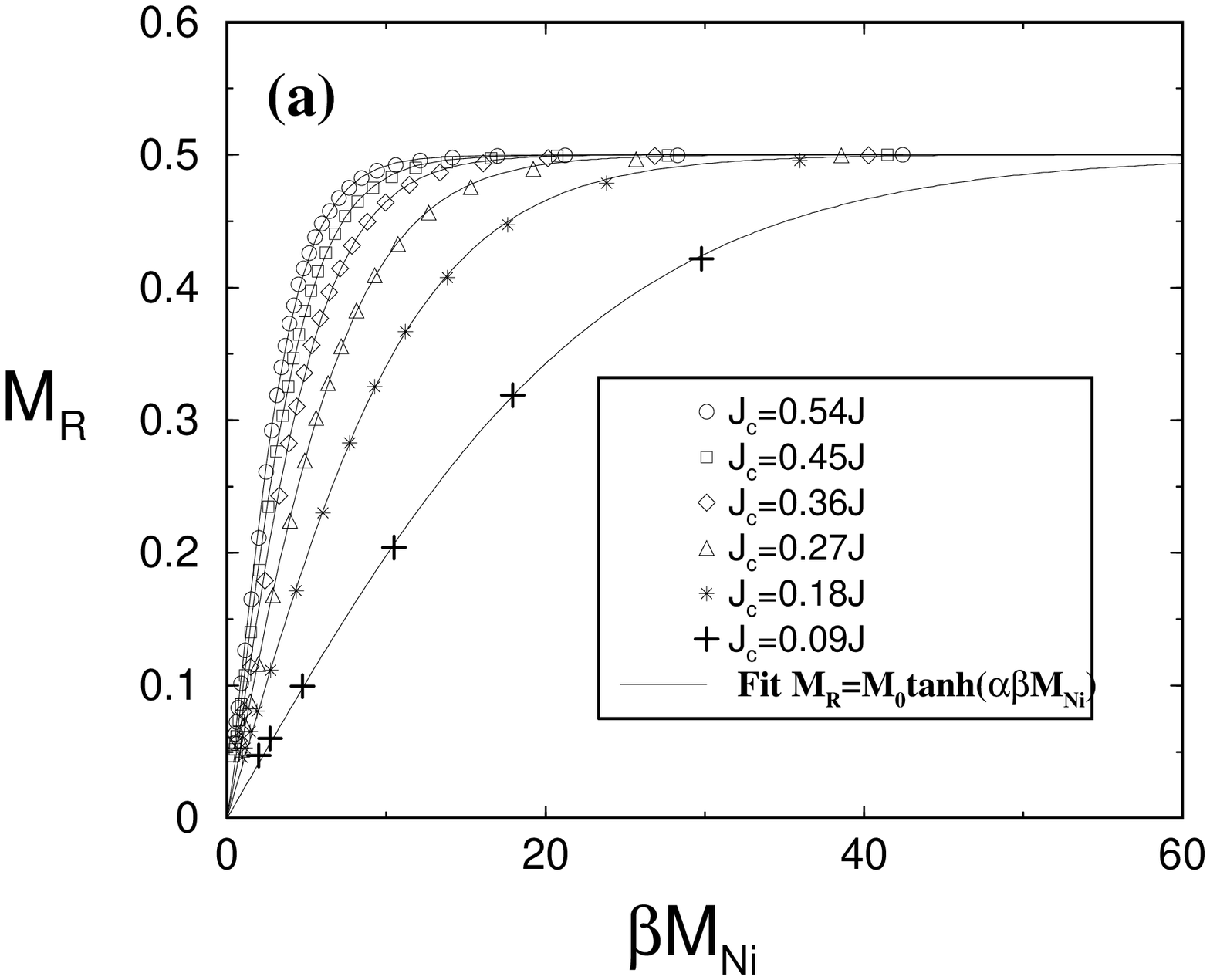}

\vspace*{0.2cm}

 \includegraphics[width=0.4\textwidth]{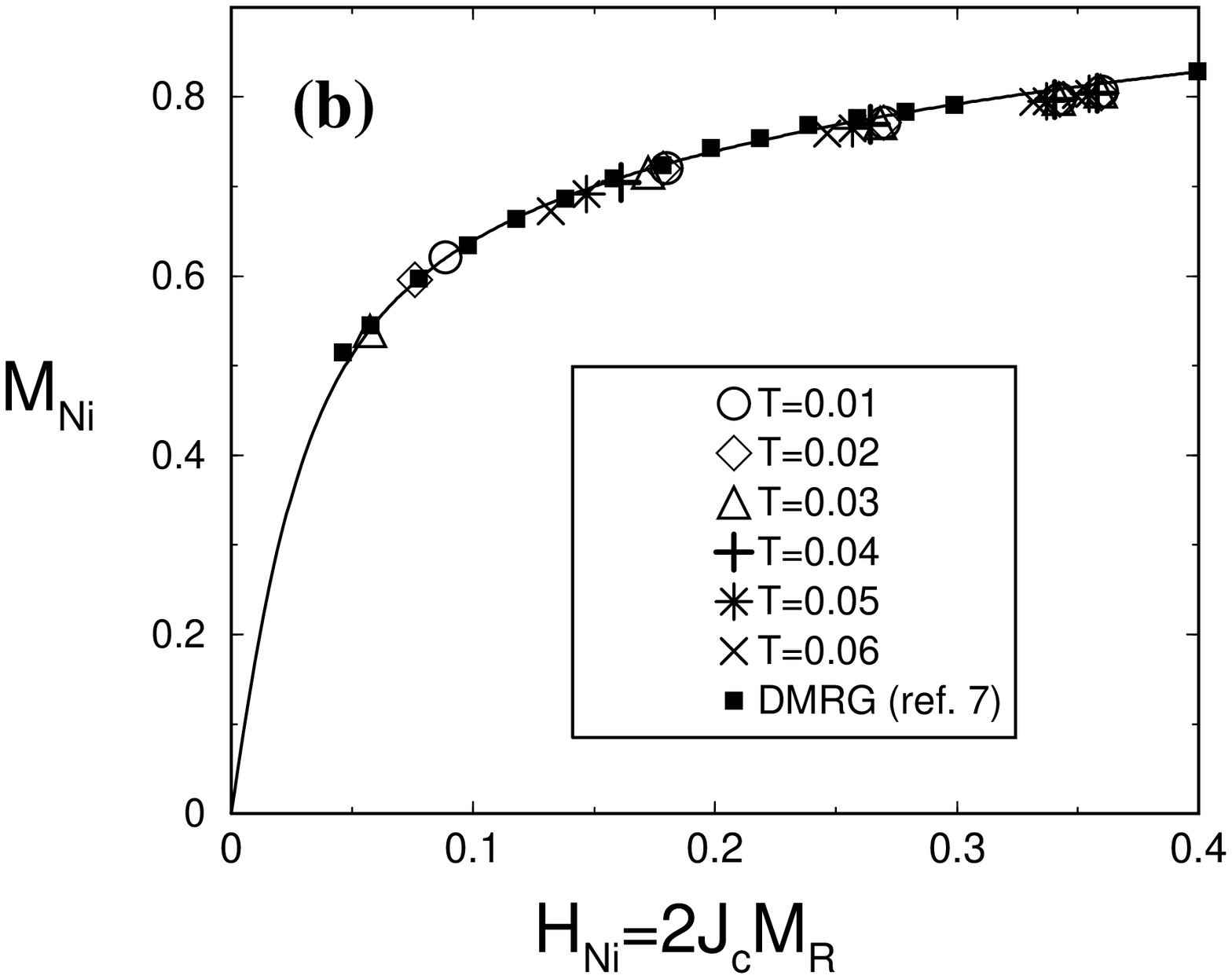}
 \caption{\label{M-H} 
QMC results for (a) $M_{\rm R}$ vs. $M_{\rm Ni}$ for various values of
the coupling constant $J_c$. The
solid lines correspond to the function Eq.\
(\protect\ref{magnetization_12}).
 (b) $M_{\rm Ni}$ vs. the
'effective' field on the Ni-subsystem induced by $M_{\rm R}$
\protect\cite{ZHELUDEV98} for various T values. The squares
correspond to the DMRG results\protect\cite{YU99} at T=0.}
 \end{figure}

In order to understand the origin of the ordered phase of this
quantum model we have studied the relation between the
magnetizations in both sublattices for different values of $J_c$.
In Fig.\ (\ref{M-H})(a) we show a plot of the QMC results at $T <
T_N$ for  $M_{R}$ as a function of  $M_{Ni}$  for various values
of $J_c$. We observe that the behavior of the magnetizations  is
well reproduced by the function:
\begin{eqnarray}
M_{\rm R}= M_0 \tanh (\alpha \beta M_{\rm Ni})
\label{magnetization_12}
\end{eqnarray}
where $\beta = 1/k_B T$, $M_0$ is the effective moment of the rare
earth ion and $\alpha$ is a linear function of $J_c$ in the range
$0.09 < J_c < 0.5J$ as observed from our QMC results.
 Eq.\ (\ref{magnetization_12}) can be obtained by
a mean-field approach as shown by Zheludev {\it et
al.}\cite{ZHELUDEV98}. These authors considered that the behavior
of the  R$_{2}$BaNiO$_{5}$ compounds in the ordered phase could be
described in terms of a $S=1$ chain of Ni ions in a staggered
magnetic field induced by the $s=\frac{1}{2}$ magnetic rare earth
ions and, reciprocally, the otherwise free spins $s=\frac{1}{2}$
see the mean field produced by the neighboring $S=1$ chains.


 The
excellent agreement between this mean-field approach and our QMC
results indicates that our microscopic model also fulfills the
property of viewing the effect of the staggered magnetization on
the rare earth ions as an effective magnetic field on the Ni
subsystem and viceversa.

 In Fig.\ (\ref{M-H})(b) we show the relation
between $M_{\rm Ni}$ and the staggered magnetic field induced by
$M_{\rm R}$ obtained by our QMC calculations for various T values.
The extrapolation $T \rightarrow 0$ reproduces the DMRG  results
for one $S=1$ chain in a staggered magnetic field at $T=0$
obtained by Yu {\it et al.}\cite{YU99}. Since the R sublattice is
fully polarized at $T=0$ no rescaling is necessary to compare our
data with the DMRG results. The quality of the fits in  Fig.\
(\ref{M-H})(a) and the data collapse in Fig.\ (\ref{M-H})(b)
suggests a technical procedure to extrapolate QMC data down to
zero temperature. This procedure is obvious for magnetizations but
could be extended for other thermodynamic magnitudes for
Hamiltonians similar to the one proposed here.

      In the context of the effective field interpretation, an
analysis of the  temperature dependence of $M_{\rm R}$  obtained
with our QMC simulations indicates that though $M_{\rm R}$
saturates to the classical value at T=0 (see Fig.\ (\ref{M-T})),
as expected if we consider this subsystem  as a sublattice of free
spins in the presence of an external field, this behavior cannot
be reproduced by a purely classical mean field model in the R
sublattice, namely mean field just proportional to the
magnetization in the R sublattice and independent of the Ni
magnetization. This fact is signaling that quantum fluctuations in
the Ni sublattice are essential to describe the magnetization in
the R sublattice.


 \begin{figure}
 \includegraphics[width=0.35\textwidth]{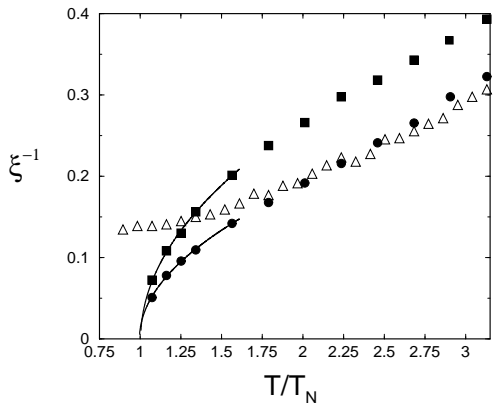}
 \caption{\label{xi-T}
Inverse correlation length vs. temperature for the Ni-chain
subsystem (circles), the R-subsystem along $a$ (squares) and for a
single $S=1$ chain (triangles). Solid lines correspond to the law
$\xi^{-1} \sim(T-T_N)^{\frac{1}{2}}$. At higher T, the
Ni-subsystem behaves like a Haldane $S=1$ chain.}
 \end{figure}


We discuss now the spatial  spin-spin correlations in the
paramagnetic phase. In Fig.\ (\ref{xi-T}) we plot  the inverse of
the correlation length $\xi^{-1}$ along the Ni chains (circles)
and along the R chains  parallel to the Ni chains (squares) for
$T<\frac{\Delta}{2}$. For comparison, we present $\xi^{-1}$ in a
single $S=1$ spin chain (triangles). In this range of temperatures
it is well known that the correlation length in the $S=1$ chain is
$\xi=\frac{\hbar c}{\Delta}$ where $c$ is the spin-excitation
velocity. The correlation length is computed in all the cases by
fitting the z-component of the spin-spin correlation function to
the law\cite{WHITE93}:
\begin{eqnarray}
|\langle S^{z}_0 S^{z}_l \rangle| =
       A\exp(\frac{-l}{\xi})l^{-\eta}
\label{eta}
\end{eqnarray} 
($A$ and $\eta$ are fitting parameters) directly in a lattice of
64$\times$64 spins and for $J_{c}=0.18$ to keep $T_{N}$ and
$\Delta$ well separated in energy. Near  $T_N$, the critical modes
become gapless and we observe how the gap closes in both
subsystems following a law $\xi^{-1} =K(T-T_N)^{\frac{1}{2}}$. K
is the only free parameter in the fit presented in Fig.\
(\ref{xi-T}) as a solid line, since  $T_N$ is computed
independently. As the temperature increases, the correlation
length in the Ni subsystem approaches the correlation length of a
single chain.

\begin{figure}
 \includegraphics[width=0.35\textwidth]{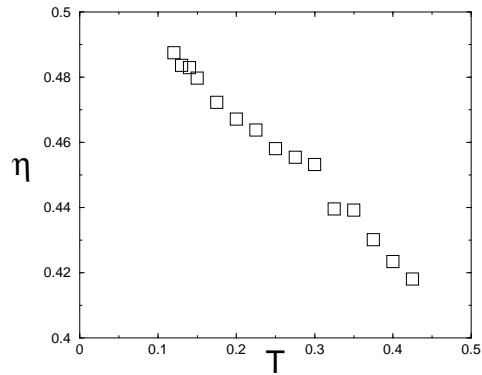}
 \caption{\label{FIG5}
 Temperature dependence of the fitting parameter $\eta (T)$ in Eq.
(\protect\ref{eta}) for the Ni-subsystem
  .}
 \end{figure}

The behavior of $\eta$ is completely different in both
sublattices. While in the  R sublattice the correlation function
$\langle S^{z}_0 S^{z}_l \rangle$
 can be fitted with $\eta=0$ for all the
temperatures showed in Fig.\ (\ref{xi-T}), in the Ni sublattice
$\eta$ approaches the value  $0.5$ in Eq.\ (\ref{eta})
as the temperature is reduced
in the paramagnetic phase. We have analized this behavior and
observed that the  temperature dependence of $\eta(T)$ in the 
spin-spin correlation function Eq.\ (\ref{eta})  is
 very similar to that
of $\eta$ in 
a single chain in which Eq.\ (\ref{eta}) shows a slow cross-over 
from $\eta=\frac{1}{2}$ at T=0, more precisely 
$\langle S^{z}_0 S^{z}_l \rangle$ is then described by the modified Bessel
function $K_{0}$ \cite{SORENSEN}, to $\eta=0$  when $T > \Delta$ \cite{KIM00}. 
The temperature dependence of $\eta(T)$ is shown in Fig.\ (\ref{FIG5}).


The picture arising from the study of the spin-spin correlation
functions is that the modes becoming gapless at $T_N$ can be
described at a mean field level. Moreover, since both  the value
of the correlation length and the power law corrections to the
exponential behavior are very similar for the Ni subsytem and for
an independent $S=1$ chain, we can conclude that the modes
remaining gapped in the system are very much like the conventional
Haldane excitations.

In summary, we have shown that QMC calculations based on the
proposed microscopic model are able to reproduce the observed
behavior of R$_2$BaNiO$_5$ mixed-spin quantum antiferromagnets
remarkably well. They provide a solid numerical basis for the
MF/RPA model, and make additional predictions regarding the
magnetic nature of both spin species that go beyond the simplistic
picture of a MF/RPA model. We hope that the proposed approach will
be particularly useful in the study of dilute
(R$_x$Y$_{1-x}$)$_2$BaNiO$_5$ compounds in the limit of small $x$,
where the MF model becomes totally inadequate.

       It is a pleasure for us to acknowledge discussions with C.
       Gros, F. Mila and H. Rieger. Oak Ridge National Laboratory
       is managed by UT-Battelle, LLC for the U.S. Department of
       Energy under contract DE-AC05-00OR22725.

*Previous address:
 Physics Department\ , Brookhaven National Laboratory, Upton,
NY 11973-5000, USA.

\end{document}